\documentclass[]{heart2015_WN4Pre}

\usepackage{listings}
\usepackage{color}
\usepackage{graphicx}
\usepackage[export]{adjustbox}
\usepackage{amsmath}
\usepackage{flushend}
\title{An Intermediate Language and Estimator for Automated Design Space Exploration on FPGAs}
\numberofauthors{2}
\author{ 
	\alignauthor Syed Waqar Nabi\\
	\affaddr {School of Computing Science,}\\
	\affaddr {University of Glasgow,} \\
	\affaddr {Glasgow G12 8QQ, UK.}\\ 
	\email {syed.nabi@glasgow.ac.uk}
	\alignauthor Wim Vanderbauwhede\\
	\affaddr {School of Computing Science,}\\
	\affaddr {University of Glasgow,} \\
	\affaddr {Glasgow G12 8QQ, UK.}\\ 
	\email {wim.vanderbauwhede@glasgow.ac.uk}\\	
}

\begin{document}

\lstset{ %
  backgroundcolor=\color{white},   
  basicstyle=\scriptsize\ttfamily,
  breakatwhitespace=false,         
  breaklines=true,                 
  captionpos=b,                    
  commentstyle=\color{black},    
  deletekeywords={...},            
  frame=single,                   
  keepspaces=true,                 
  keywordstyle=\color{blue},       
  language=Octave,                 
  morekeywords={*,...},            
  numbers=left,                    
  numbersep=5pt,                   
  numberstyle=\tiny\color{black}, 
  rulecolor=\color{black},         
  showspaces=false,                
  showstringspaces=false,          
  showtabs=false,                  
  stepnumber=1,                    
  tabsize=2,                       
}

\maketitle

\begin{abstract}
We present the TyTra-IR, a new intermediate language intended as a compilation target for high-level language compilers and a front-end for HDL code generators. 
We develop the requirements of this new language based on the \textit{design-space} of FPGAs that it should be able to express and the \textit{estimation-space} in which each configuration from the design-space should be mappable in an automated design flow. We use a simple kernel to illustrate multiple configurations using the semantics of TyTra-IR. The key novelty of this work is the cost model for resource-costs and throughput for different configurations of interest for a particular kernel. Through the realistic example of a Successive Over-Relaxation kernel  implemented both in TyTra-IR and HDL, we demonstrate both the expressiveness of the IR and  the accuracy of our cost model.
\end{abstract}

\section{Introduction}
\label{sec:intro}

The context for  the work in this paper is the \textit{TyTra} project \cite{tytra_website}which aims to develop  a compiler for heterogeneous platforms for high-performance computing (HPC) that includes many/multi-core CPUs, graphics processors (GPUs) and Field Programmable Gate Arrays (FGPAs). The work we present here relates to raising the programming abstraction for targeting FPGAs, reasoning about its multi-dimensional  design space, and estimating parameters of interest of multiple configurations from this design-space via a cost model. 

We present a new language, the \textit{TyTra Intermediate Representation} (TIR) language, which has an abstraction level and syntax intentionally similar to the LLVM Intermediate Language \cite{123.303}. We can derive resource-utilization and performance estimates from TIR code via a light-weight back-end compiler, \textit{TyBEC}, which will also generate the HDL code for the FPGA synthesis tools. We will briefly discuss syntax of the IR and its expressiveness through illustrations, and discuss the cost model we have built around this language.

The TyTra project is predicated on the observation that we have entered a period where performance increases can only come from increased numbers of heterogeneous computational cores and their effective exploitation by software. 
%
%
The specific challenge we are addressing in the TyTra project is how to exploit the parallelism of a given computing platform, e.g. a multicore CPU, GPU or a FPGA, in the best possible way, without having to change the original program.
Our proposed approach is to use an advanced type system called Multi-Party Session Types \cite{123.203} to describe the communication between the tasks that make up a computation, to transform the program using provably correct type transformations, and to use machine learning and a cost model to select the variant of the program best suited to the heterogeneous platform.  Our proof-of-concept compiler is being developed and targets FPGA devices, because this type of computing platform is the most different from other platforms and hence the most challenging.

\begin{figure}[th]
\centering
\includegraphics[width=0.7\linewidth]{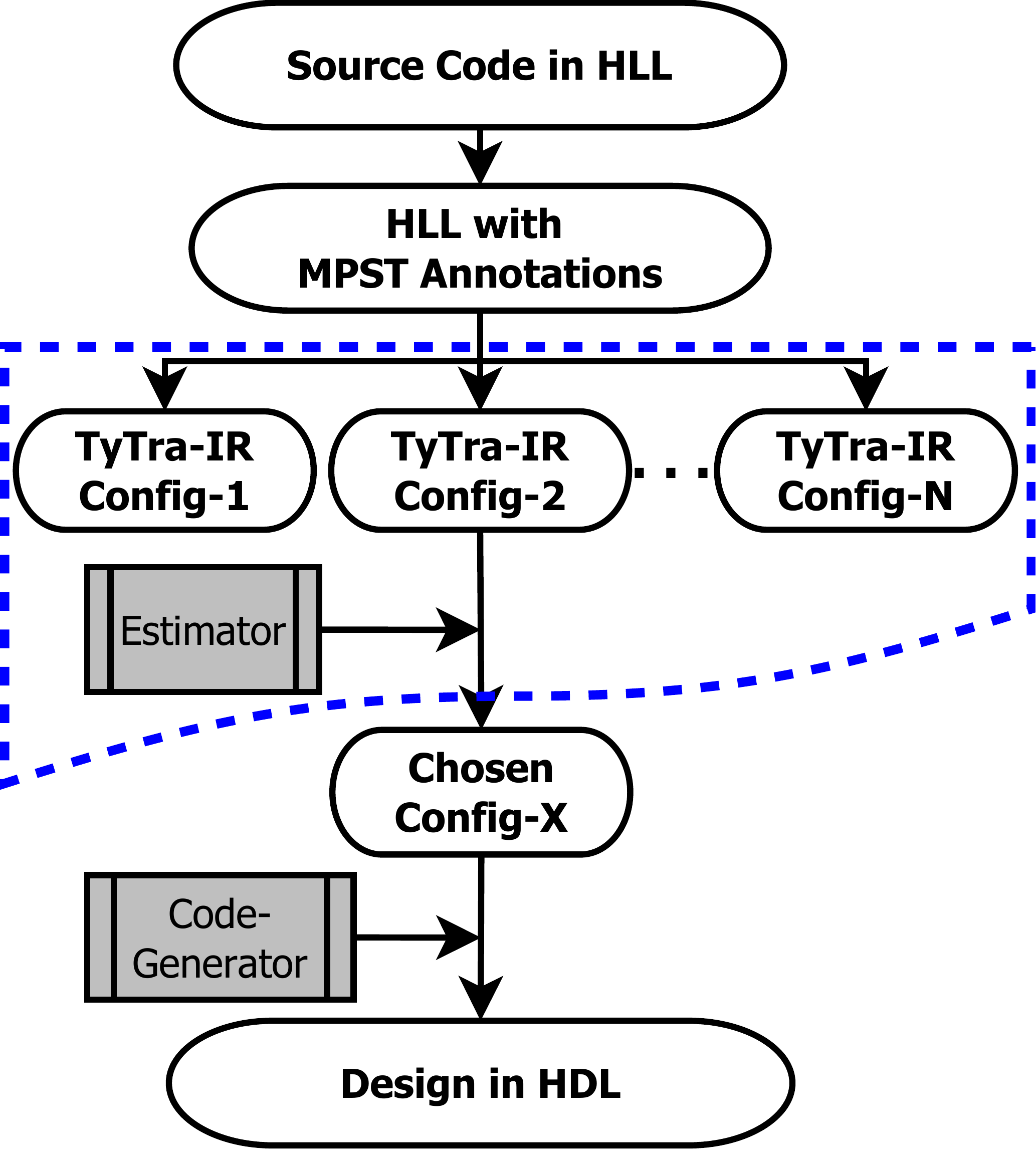}
\caption{The TyTra project design flow. This paper focuses on the area marked out by dotted lines.}
\label{fig:TyTra_Flow}
\end{figure}

Figure \ref{fig:TyTra_Flow} is a concise representation of the compiler's expected flow. The work we present in this paper --- identified by the dotted box --- is limited to the specification and abstraction of the IR, its utility in representing various configurations, and the cost model built around it which we can use to assess the trade-offs associated with these configuration. 

In the next section, we present the TyTra platform model for FPGAs. A very important abstraction for this work is our view of the \textit{design-space} of FPGAs, which we present next, followed by something we call the \textit{estimation-space}. Both the design-space and estimation-space are our attempts to give structure to reasoning around multiple configurations of a kernel on an FPGA. We then specify the requirements of the new IR language that we are developing. We follow this with a very brief description of the TIR, and develop this by a expressing different FPGA configurations. We then look at the scheme to arrive at various estimates, and evaluate it by a simple example based on the successive-relaxation method. We conclude by briefly discussing some related work and our future lines of investigation.
 
\section{Platform Model}
\label{sec:platform-model}

\begin{figure}
\centering
\includegraphics[width=1.0\linewidth]{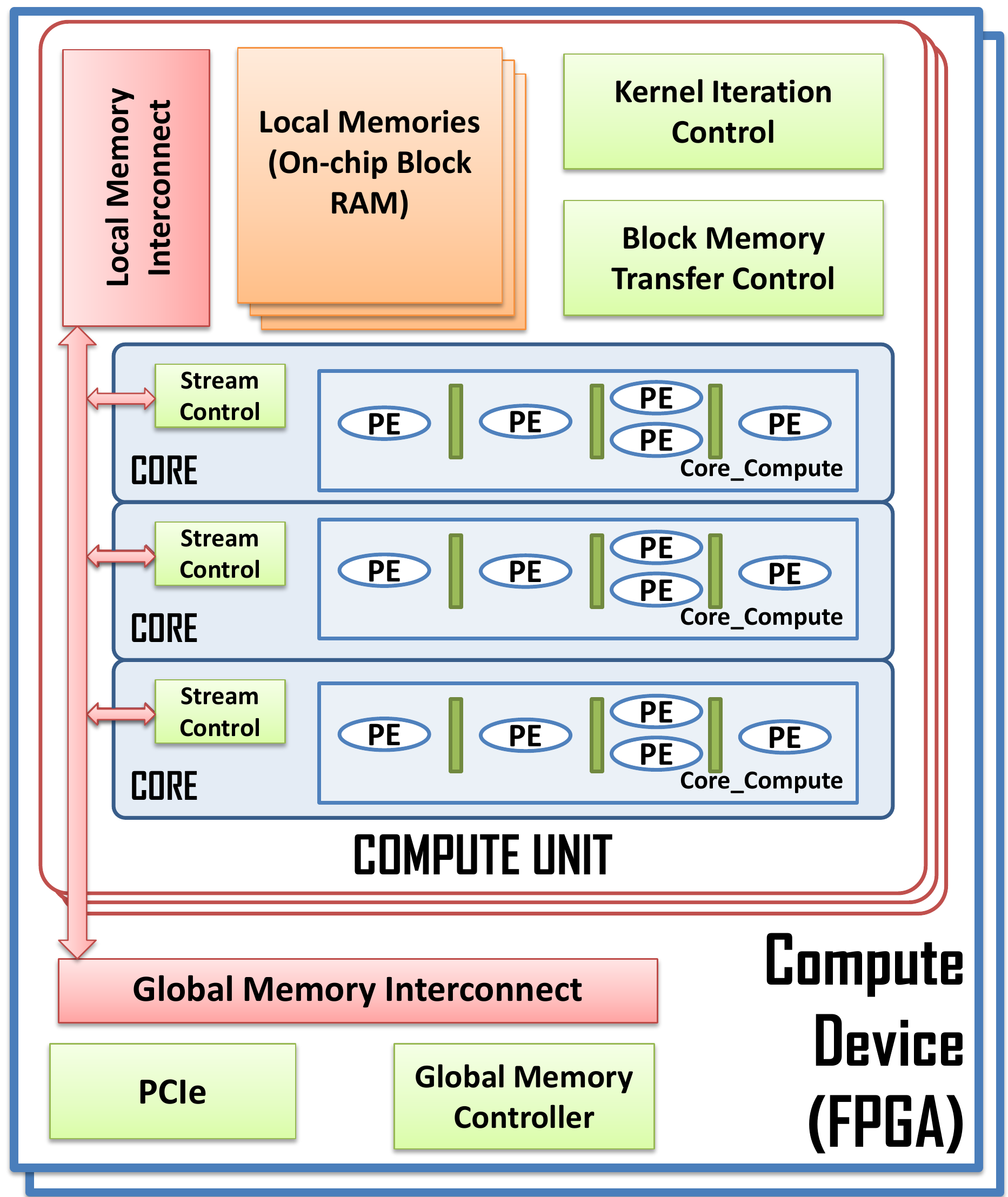}
\caption{The TyTra-FPGA Platform Model, showing a typical pipelined \texttt{Core\_Compute} incorporating ILP, and replicated for thread-parallelism.}
\label{fig:TytraPlatformModel4Heart15}
\end{figure}

The Tytra-FPGA platform model is similar to the  platform model introduced by OpenCL \cite{112.155}, but also informed by  our prior work on the MORA FPGA  programming framework \cite{chalamalasetti2009mora}, and more nuanced than OpenCL's to incorporate FPGA-specific architectural features; Altera-OpenCL takes a similar approach \cite{112.151}. The main departure from the OpenCL model is the \textit{Core} block, and the \textit{Compute-Cores}. Figure \ref{fig:TytraPlatformModel4Heart15} is a block diagram of the model, with brief description following. We do however use the terms global memory, local memory, work-group, and work-item, exactly as they are used in the OpenCL framework.

\begin{description}
	\item [{Compute-Device}] An FPGA device, which would contain one or more \textit{compute-units}. 
	
	\item [{Compute-Unit}] Execution unit for  a single kernel. An FPGA allows multiple independent kernels to be executed concurrently, though typically there would be a single kernel. The compute-unit contains local memory (block RAM), some custom logic for controlling iterations of a kernel's execution and managing block memory transfers, and one or more \textit{cores}.
	
	\item [{Core}] This is the custom design unit created for a kernel. For pipelined implementations, a core may be considered equivalent to a pipeline \textit{lane}. There can be multiple lanes for thread-level parallelism (TLP). The core has control logic for generating data streams from a variety of data sources like local-memory, global-memory, host, or a peer compute-device or compute-unit. These streams are fed to/from the \textit{core-compute} unit inside it, which consists of \textit{processing-elements (PEs)}. A PE can consist of an arithmetic or logic functional unit and its pipeline register, or it can also be a custom scalar or vector instruction processor with its own \textit{private memory}.
\end{description}

\section{Configuration, Performance and Cost Abstractions}
\label{sec:design-estimation-space}

As FPGAs have a fine-grained flexibility, parallelism in the kernel can be exposed by different configurations.  It is useful to have some kind of a structure to reason about these configurations; much more so when the goal is an automated design flow. We have created a design-space abstraction for the key differentiating feature of concern of multiple FPGA configurations --- the kind and extent of parallelism available in the design. We define an \textit{estimation-space} for capturing the various estimates for a point in the design-space. By defining a design-space, an estimation-space, and a mapping between them, we have a structured approach for mapping a particular kernel to a suitable configuration on the FPGA.

\subsubsection*{Design Space}

\begin{figure}[thb]
\centering
\includegraphics[width=1.0\linewidth]{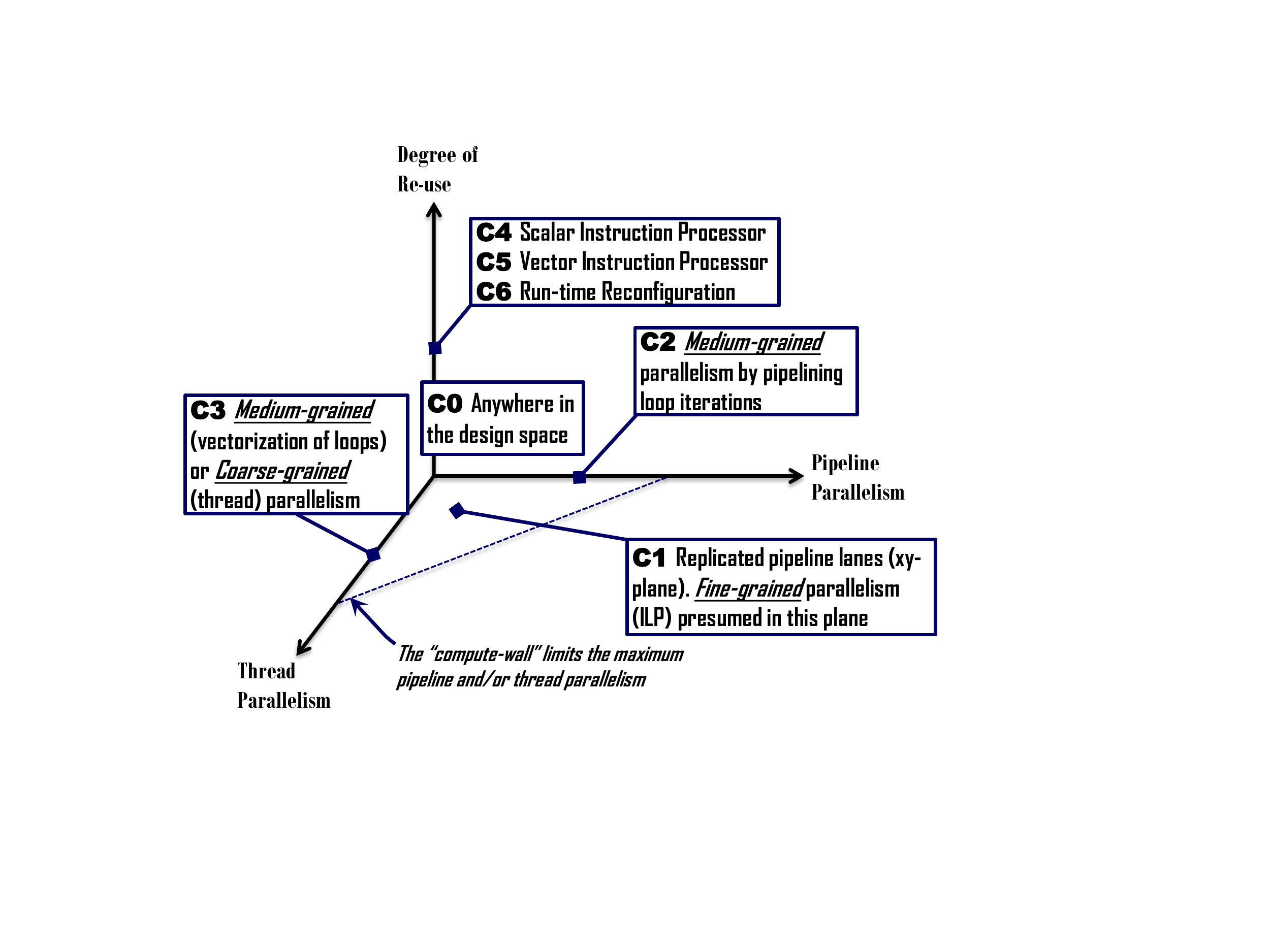}
\caption{The TyTra-FPGA Design Space Abstraction}
\label{fig:designSpace}
\end{figure}

The \textit{design-space} is shown in Figure \ref{fig:designSpace}. A \texttt{C2} configuration, on the axis indicating the degree of pipeline parallelism, is a pipelined implementation of the kernel on the FPGA. The other horizontal axis indicates the degree of parallelism achieved by replicating a kernel's core. This can be done by simultaneously launching multiple calls to a kernel, which is parallelism at a coarse, thread level. Along the same dimension is a medium-grained parallelism, which involves launching multiple work-items of a kernel's work-group. 

A configuration in the xy-plane, \texttt{C1}, has multiple kernel cores, each of which has pipeline parallelism as well. We expect this to be the preferable configuration for most small to medium sized kernels, where the FPGA has enough resources to allow multiple kernel instantiations to reside simultaneously. 

Note that we have not explicitly shown the most fine-grained parallelism, i.e., Instruction-Level Parallelism (ILP). The assumption is that it will be exploited wherever possible in the pipeline.

While our current focus is on kernels where we can fit at least one fully laid out custom pipeline on the available FPGA resources, re-use of logic resources is possible for larger kernels by cycling through some instructions in a scalar (\texttt{C4}) or vector (\texttt{C5}) fashion, or by using the run-time reconfiguration capabilities of FGPA devices to load in and out relevant subsets of the kernel implementation (\texttt{C6}). %

Finally, \texttt{C0} represents the generic configuration for any point on the design space. 

\subsubsection*{Estimation Space}

\begin{figure}
\centering
\includegraphics[width=1.0\linewidth]{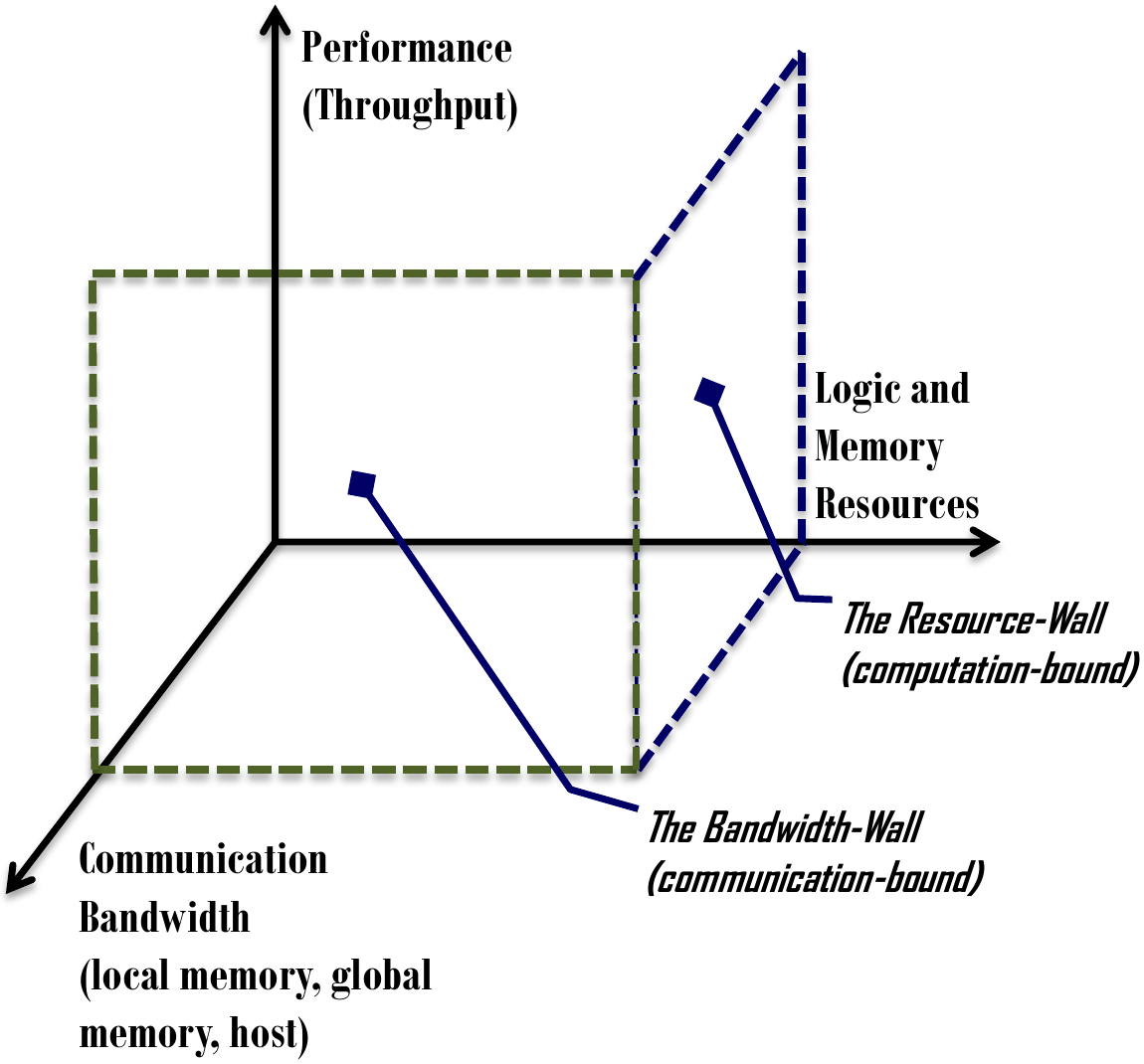}
\caption{The TyTra-FPGA Estimaton Space Abstraction}
\label{fig:TyTraCostSpace4Heart15}
\end{figure}

The TyTra design flow (Figure \ref{fig:TyTra_Flow}) depends on the ability of the compiler to make choices about configurations from the design-space of a particular kernel on an FPGA device. Various parameters will be relevant when making this choice, and the success of the TyTra compiler is predicated on the ability to derive estimates of \textit{reasonable} accuracy for these parameters of concern from the IR, without actually having to generate HDL code and synthesize each configuration on the FPGA. The estimation-space as shown in Figure \ref{fig:TyTraCostSpace4Heart15} is a useful abstraction in this context. The obvious aim is to go as high up as possible on the performance axis, while staying within the computation and IO constraint walls. 

Having developed the design-space and estimation-space, it follows that the TyTra-IR should intrinsically be capable of working with both these abstractions, as we discuss in the next section.

\section{Requirements for TyTra-IR}
\label{sec:tir-requirements}


The TyTra-IR is one of the key  technologies in our proposed approach, hence its design needs to meet many requirements:

\begin{enumerate}
	\item Should be intrinsically expressive enough to explore the entire design space of an FPGA (Figure \ref{fig:designSpace}), but with a particular focus on custom pipelines because our prime target is HPC applications\cite{112.152}. (The \texttt{C1} plane).

	\item Should make a convenient target for a front-end compiler that would emit multiple versions of the IR (See Figure \ref{fig:TyTra_Flow}).
	
	\item Should be able to express access operations in the entire communication hierarchy of the target device\footnote{We have omitted the details in this paper, but the TyTra memory-model extends that of LLVM.}.
	
	\item Should allow custom number representations to fully utilize the flexibility of FPGAs. If this flexibility offered by FPGAs is not capitalized on, it will be difficult to compete with GPUs for use in HPC for most scientific applications \cite{130.w1a_02}.
	
	\item The language should have enough detail of the architecture to allow generation of synthesizeable HDL code.
	
	\item A core requirement is to have a light-weight cost-model for high-level estimates. We should be able to place each configuration of interest in the design space (Figure \ref{fig:designSpace}) to a point in estimation space (Figure \ref{fig:TyTraCostSpace4Heart15}).
	
\end{enumerate}



The above requirements necessitate the design of a custom intermediate language, as none of the existing HLS ("C-to-gates") tools meets all requirements. HLS front-end languages are primarily focused on ease of human-use. High-level hardware description languages like OpenCL or MaxJ\cite{123.024}, having coarse-grained high-level datapath and control instructions and syntactic sugar, are inappropriate as compiler targets because the abstraction level is too high. Moreover, even parallelism friendly high-level languages tend to be constrained to specific types of parallelism, and exploring the entire FPGA design-space would either be impossible, or protracted. The requirements of a lightweight cost-model also motivated us to work on a new language.

\section{The Tytra-IR}
\label{sec:TIR}

The TyTra-IR (TIR) is a strongly and statically typed language, and all computations are expressed using Static Single Assignments (SSA). %
The TIR is largely based on the LLVM-IR because it gives us a suitable point of departure for designing our language, where we can re-use the syntax of the LLVM-IR with little or no modification, and allows to explore LLVM optimizations to improve the code generation capabilities of our tool, as e.g. the LegUp \cite{112.153} tool does. 
%
%
%
We use LLVM metadata syntax and some custom syntax as an abstraction for FPGA-specific architectural features. 

The TIR code for a design has two components:
\begin{description}
	\item [{Manage-IR}] deals with setting	up the streaming data ports for the kernel. It corresponds to the logic in the \textit{core} outside the \textit{core-compute} (See Figure \ref{fig:TytraPlatformModel4Heart15}). All Manage-IR statements are wrapped inside the \texttt{launch()} method.
	
	\item [{Compute-IR}] describes the datapath logic that maps to the core-compute unit inside the core. It mostly works with very limited data abstractions, namely, streaming and scalar ports. All Compute-IR statements are in the scope of the \texttt{main()} function or other functions ``called'' from it.
\end{description}

By dividing the IR this way, we separate the pure dataflow architecture --- working with streaming variables and arithmetic datapath units ---  from the control and peripheral logic that creates these streams and related memory objects, instantiates required peripherals for the kernel application, and manages the host, peer-device, and peer-unit interfaces. The division between compute-IR and manage-IR directly relates to the division between core-compute unit and the remaining core logic (wrapper) around it (See Figure \ref{fig:TytraPlatformModel4Heart15}). A detailed discussion of the TIR syntax is outside the scope of this paper, but the following illustration of its use in various configurations gives a good picture.

\section{Illustration of IR Use}
\label{sec:TIR-illustration}

We use a trivial example and build various configurations for it, to demonstrate the architectural expressiveness of the TIR. The following Fortran loop describes the kernel:
\begin{verbatim}
do n = 1,ntot
  y(n) =  K + ( (a(n)+b(n)) * (c(n)+c(n)) )                
end do 
\end{verbatim}

\subsection{Sequential Processing}
\label{sec:illustration-config1}

\begin{figure}[tbh]
\centering
\includegraphics[width=0.9\linewidth, frame]{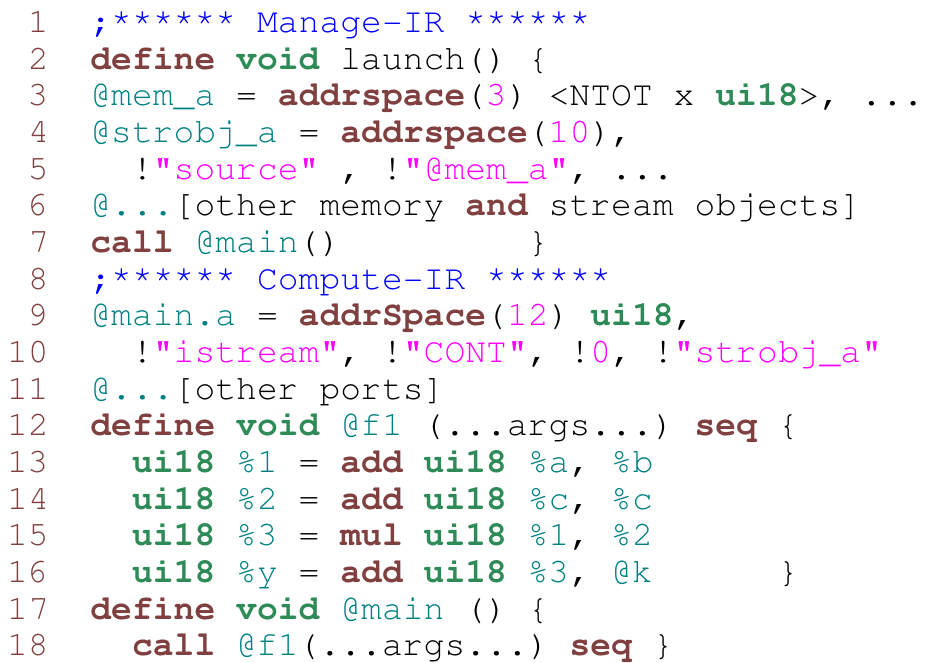}
\caption{TyTra-IR code for a sequential processing configuration of a simple kernel}
\label{fig:code4paperC4}
\end{figure}

The baseline configuration, whose redacted TIR code is showed in Figure \ref{fig:code4paperC4}, is simply a sequential processing of all the operations in the loop. This corresponds to \texttt{C4} configuration in Figure \ref{fig:designSpace}.

The manage-IR consists of the launch method which sets up the \textit{memory-objects}, which are abstractions for any object that can be the source or destination of streaming data. In this case, the memory object (Figure \ref{fig:code4paperC4}, line 3) is a local-memory instance, indicated by the argument to \texttt{addrspace} qualifier. The stream-objects connect to memory-objects to create streams of data, as shown in lines 4--5. The creation of streams from memory is equivalent reading from an array in a loop, hence we see that the loop over work-items in Fortran disappears in the TIR. After setting up all stream and memory objects, the main function is called. 

\begin{figure}[th]
\centering
\includegraphics[width=0.8\linewidth]{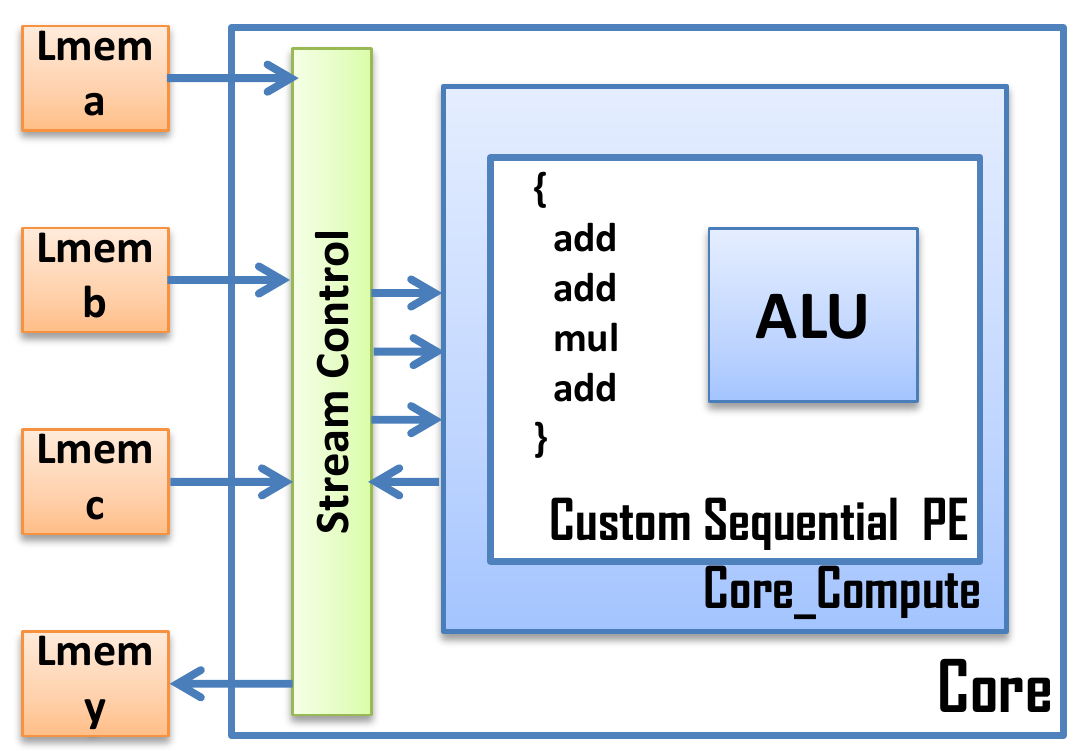}
\caption[]{Sequential configuration}
\label{fig:config3}
\end{figure}

The compute-IR sets up the ports (lines 9-11), which are mapped to a stream-object, creating data streams for the compute-IR functions. The SSA datapath instructions in function \texttt{f1} are configured for sequential execution on the FPGA, indicated by the keyword \texttt{seq}, and then \texttt{f1} is called by \texttt{main}. Figure \ref{fig:config3} shows the block diagram for this configuration.

\subsection{Single Kernel Execution Pipeline}
\label{sec:illustration-config2}

\begin{figure}[th]
\centering
\includegraphics[width=1.0\linewidth, frame]{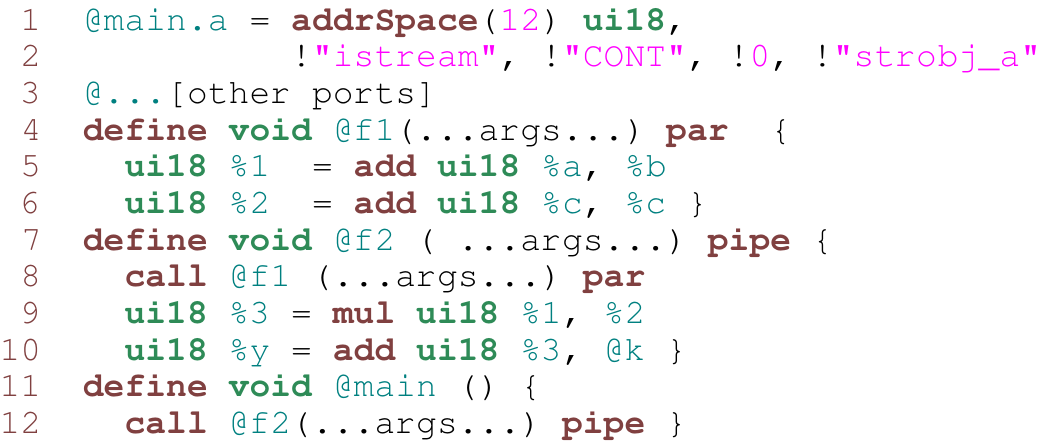}
\caption{TyTra-IR code for a pipelined configuration of a simple kernel}
\label{fig:code4paperC2}
\end{figure}

This configuration (\texttt{C2}) is a fully pipelined version of the kernel, and the TIR code is shown in Figure \ref{fig:code4paperC2}. 

\begin{figure}[th]
\centering
\includegraphics[width=0.8\linewidth]{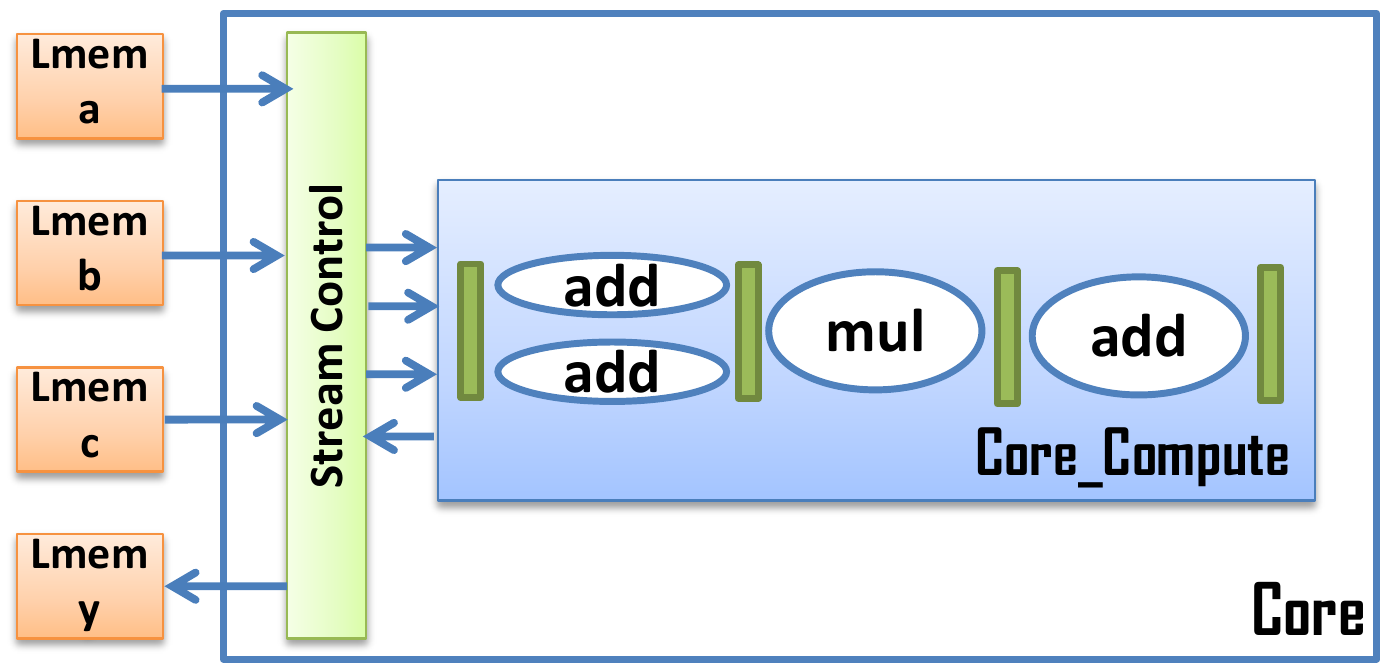}
\caption[]{Single Pipeline with ILP}
\label{fig:config1}
\end{figure}

Note that the available ILP (the two add operations can be done in parallel) is exploited by explicitly wrapping the two instructions into a \texttt{par} function \texttt{f1}, and then calling it in the pipeline function \texttt{f2}. Our prototype parser can also automatically check for dependencies in a pipe function and schedule instructions using a simple as-soon-as-possible policy. See Figure \ref{fig:config1} for the block diagram of this configuration.

\subsection{Multiple Kernel Execution Pipelines}
\label{sec:illustration-config3}

\begin{figure}[th]
\centering
\includegraphics[width=0.9\linewidth, frame]{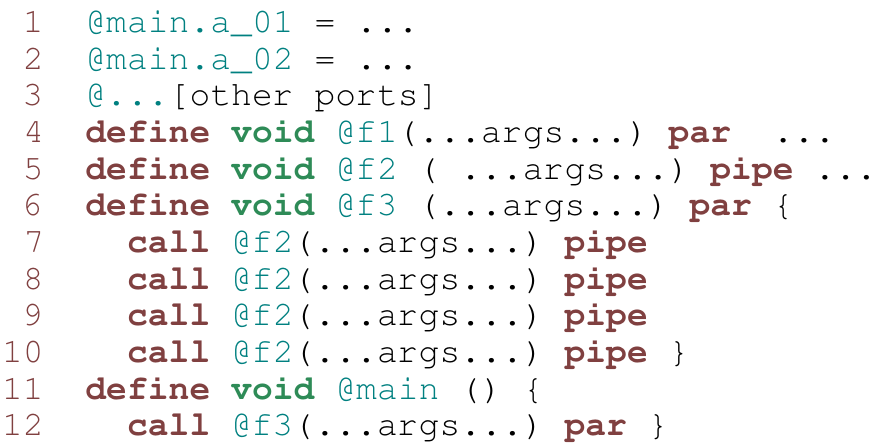}
\caption{TyTra-IR code for replicated pipeline configuration of a simple kernel}
\label{fig:code4paperC1}
\end{figure}

\begin{figure}[th]
	\centering
	\includegraphics[width=0.7\linewidth]{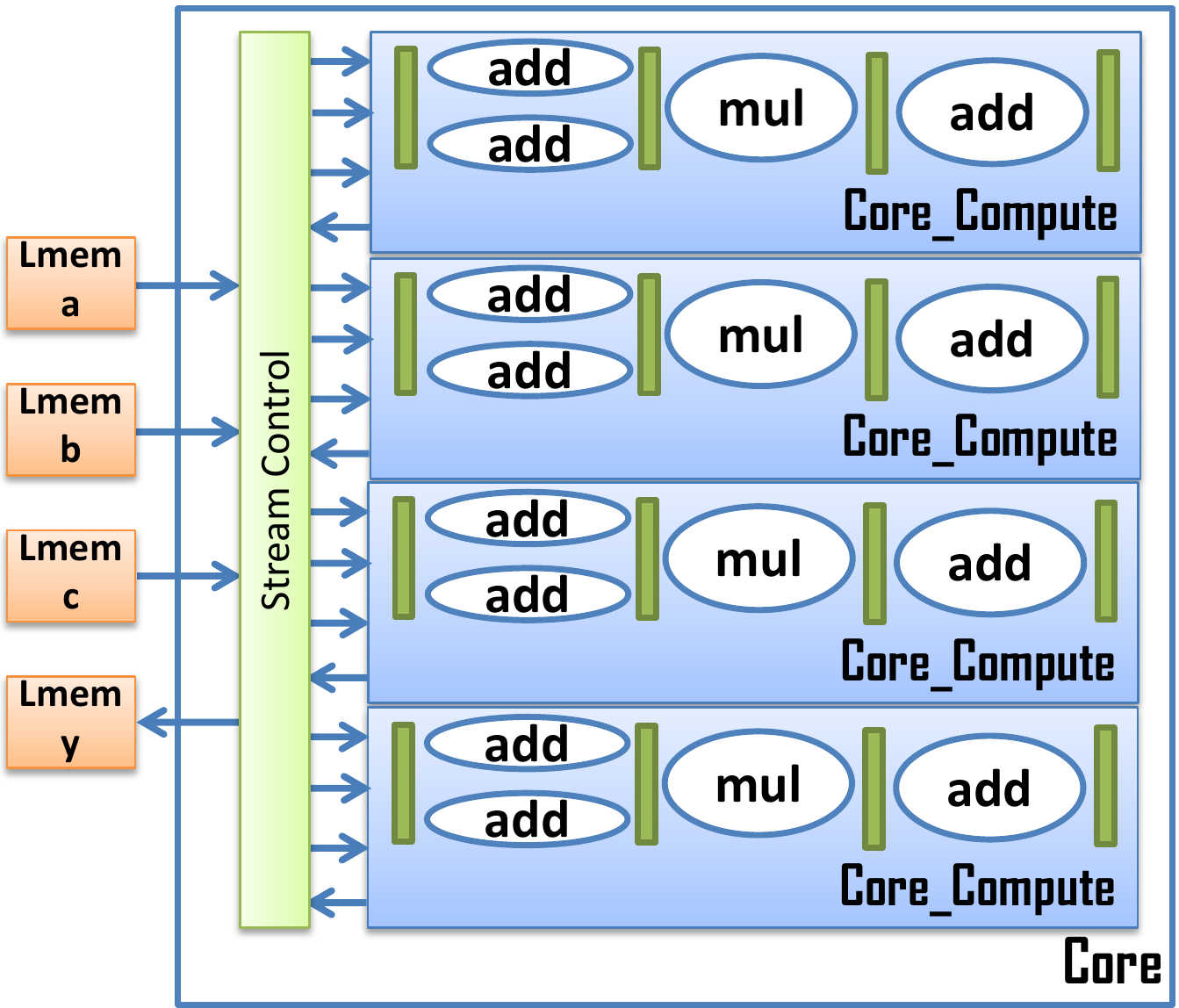}
	\caption[]{Replicated Pipelines}
	\label{fig:config2}
\end{figure}

For simple kernels where enough space is left after creating one pipeline core for its execution, we can instantiate multiple identical pipeline \textit{lanes} (\texttt{C1}). The code in Figure \ref{fig:code4paperC1} illustrates how this can be specified in TIR. We do not reproduce segments that have appeared in previous listings. See Figure \ref{fig:config2} for the block diagram of this configuration.

Comparing with the previous single-pipeline configuration, note that we have a new \texttt{par} function \texttt{f3} calling the same \texttt{pipe} function four times, indicating replication. Similarly, there are now four separate ports for each array input, and there are four separate streaming objects (not shown) for each of these ports, all of which connect to the same memory object, indicating a multi-port memory.

\subsection{Multiple Sequential Processing Elements - Vector Processing}
\label{sec:illustration-config4}

There is one more interesting configuration we can express in TIR by wrapping multiple calls to a \texttt{seq} function in a \texttt{par} function. This would represent a vectorized sequential processor (\texttt{C5}). 

\begin{figure}[th]
	\centering
	\includegraphics[width=0.9\linewidth, frame]{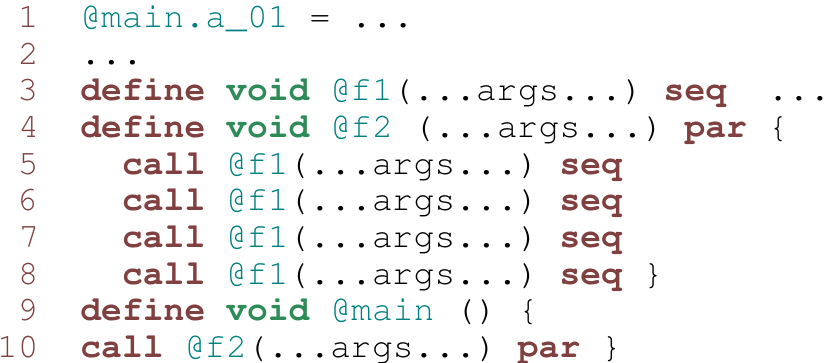}
	\caption{TyTra-IR code for vectorized sequential processing of a simple kernel}
	\label{fig:code4paperC5}
\end{figure}

The TIR for this configuration is shown in Figure \ref{fig:code4paperC5}, with only the relevant new bits emphasized. 

\begin{figure}[th]
	\centering
	\includegraphics[width=0.9\linewidth]{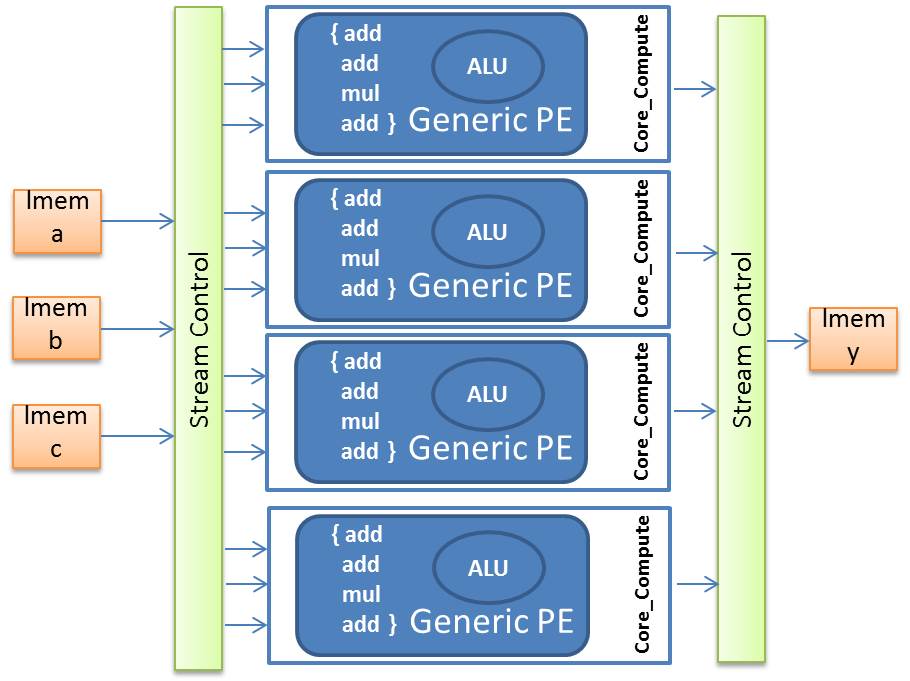}
	\caption[]{Configuration 4: Vectorized Sequential Processing}
	\label{fig:config4}
\end{figure}

See Figure \ref{fig:config4} for the block diagram of this configuration.

Comparing with the single sequential processor configuration, note that we have a new \verb|par| function \verb|f2| that calls the same \verb|seq| function four times, indicating a replication of the sequential processor.

\section{The TyTra-FPGA Cost Model} 
\label{sec:estimation-model}

Following from the requirement of the ability to get cost and performance estimates as discussed in \S\ref{sec:tir-requirements}, we  designed the TIR  specifically to allow generation of accurate estimates. Our prototype TyTra Back-end Compiler (TyBEC) can calculate estimates directly from the TIR without any further synthesis. Many different configurations for the same kernel can be compared by a programmer or -- eventually -- by a front-end compiler.

Two key estimates are calculated by the TyBEC estimator: the resource utilization for a specific Altera FPGA device (ALUTs, REGs, Block-RAM, DSPs), and the throughput estimate for the kernel under consideration. With reference to Figure \ref{fig:TyTraCostSpace4Heart15}, this covers two dimensions. An estimate of IO bandwidth requirements is on-going work. For the purpose of this work we make the simplifying assumption that all kernels are compute-bound rather than IO-bound. 

\subsection{Estimating Throughput}

We have described a performance measure called the \textbf{EWGT (Effective Work-Group Throughput)} for comparing how fast a kernel executes across different design points. This may be defined as the number of times an entire work-group (the loop over all the work-items in the index-space) of a kernel is executed every second. %
%
Measuring throughput at this granularity rather than the more conventional bits-per-second unit allows us to reason about performance at a coarse enough level to take into account parameters like dynamic reconfiguration penalty. %
Following is the generic expression which applies to the entire design space (i.e. the \texttt{C0} root configuration), and specialized expressions for configurations of interest can be derived from it:

\begin{align*}
EWGT\; & =\;\frac{L.D_{V}}{N_{R}.\left\{ T_{R}+N_{I}.N_{to}.T.\left(P+I\right)\right\} }
\end{align*}

Where:

{$EWGT=$} Effective Workgroup Throughput

{$L=$} Number of identical \emph{lanes} 

{$D_{V=}$} Degree of vectorization 

{$N_{R}=$} Number of FPGA configurations needed to execute the entire kernel 

{$T_{R}=$} Time taken to reconfigure FPGA.

{$N_{I}=$} Number of equivalent FLOP instructions delegated to the average instruction processor

{$N_{TO}=$} Ticks taken by one FLOP operation, i.e. CPI.

{$T=$} FPGA clock period.

{$P=$} Pipeline depth. 

{$I=$} Number of work-items in the kernel loop. 

The key novelty is that the TIR through its constrained syntax at a particular abstraction \textit{exposes} the parameters that make up the expression, and a simple parser can extract them from the TIR code, as we will show in \S \ref{sec:estimator-flow}. If we were to use a higher-abstraction HLS language as our internal IR representation, we would not be able to use the above expression, and some kind of heuristic would have to be involved in making the estimates.

All specialized expressions for different types of configurations can be obtained from the generic expression as follows:

For \textbf{C1}, with multiple kernel pipelines, no sequential processing, we set $N_{R}=1, T_{R}=0, N_{I}=1, D_{V}=1$, giving us: 
\begin{align*}
EWGT\; & =\;\frac{L}{N_{to}.T.\left(P+I\right)}
\end{align*}

For \textbf{C2}, limited to one pipeline lane, setting $N_{R}=1,T_{R}=0,N_{I}=1,D_{V}=1,L=1$ leads to:
\begin{align*}
EWGT\; & =\;\frac{1}{N_{to}.T.\left(P+I\right)}
\end{align*}

For \textbf{C3}, with no pipeline parallelism, we set $N_{R}=1,T_{R}=0,N_{I}=1,D_{V}=1,P=1$ to give:
\begin{align*}
EWGT\; & =\;\frac{L}{N_{to}.T.I}
\end{align*}

For \textbf{C4}, where PEs are scalar instruction processors, setting $N_{R}=1,T_{R}=0,D_{V}=1$ leads to:
\begin{align*}
EWGT\;=\;\frac{L}{N_{I}.N_{to}.T.\left(P+I\right)}
\end{align*}

For \textbf{C5}. where PEs are vector instruction processors, we set $N_{R}=1,T_{R}=0$, getting:
\begin{align*}
EWGT\;=\;\frac{L.D_{V}}{N_{I}.N_{to}.T.\left(P+I\right)}
\end{align*}

Finally, for \textbf{C6}, with multiple run-time configurations the expression remains the same as \textbf{C0}.

As an example, the single-pipelined version in \S\ref{sec:illustration-config2} corresponds to \texttt{C2}, and multi-pipeline in \S\ref{sec:illustration-config3}, corresponds to \texttt{C1}. We estimated their EWGT based on the relevant expression above, and then compared it to the figures from HDL simulation. See the comparison in the last row of Table \ref{tab:estVactualSimple}. Note that the \texttt{cycles/kernel} estimate (second-last row) is very accurate; the somewhat higher deviation of about 20\% in EWGT estimate is due to the deviation in estimation of device frequency.

\subsection{Estimating Utilization of FPGA Resources}

Each instructions can be assigned a resource cost by one of two methods:

\begin{enumerate}
	\item Use a simple analytical expression developed specifically for the device based on experiments. We have found that the regularity of FPGA fabric allows some very simple first or second order expressions to be built up for most instructions based on a few experiments. The details are outside the scope of this paper.
	\item Lookup, and possibly interpolate, from a cost database for the specific token and data type. 
\end{enumerate}

The resource costs are then accumulated based on the structural information available in the TIR. For example, two instructions in a \texttt{pipe} function will incur additional cost of pipeline registers, and instruction in a \texttt{seq} block will save some resources by re-use of functional units, but there will be an additional cost of storing the instructions, and creating control logic to sequence them on the shared functional units.
%
Both the cost and performance estimates follow trivially once we have the kernel expressed in the TIR abstraction.

\subsection{The TyTra Estimator Flow}
\label{sec:estimator-flow}

\begin{figure}[tbh]
\centering
\includegraphics[width=0.7\linewidth]{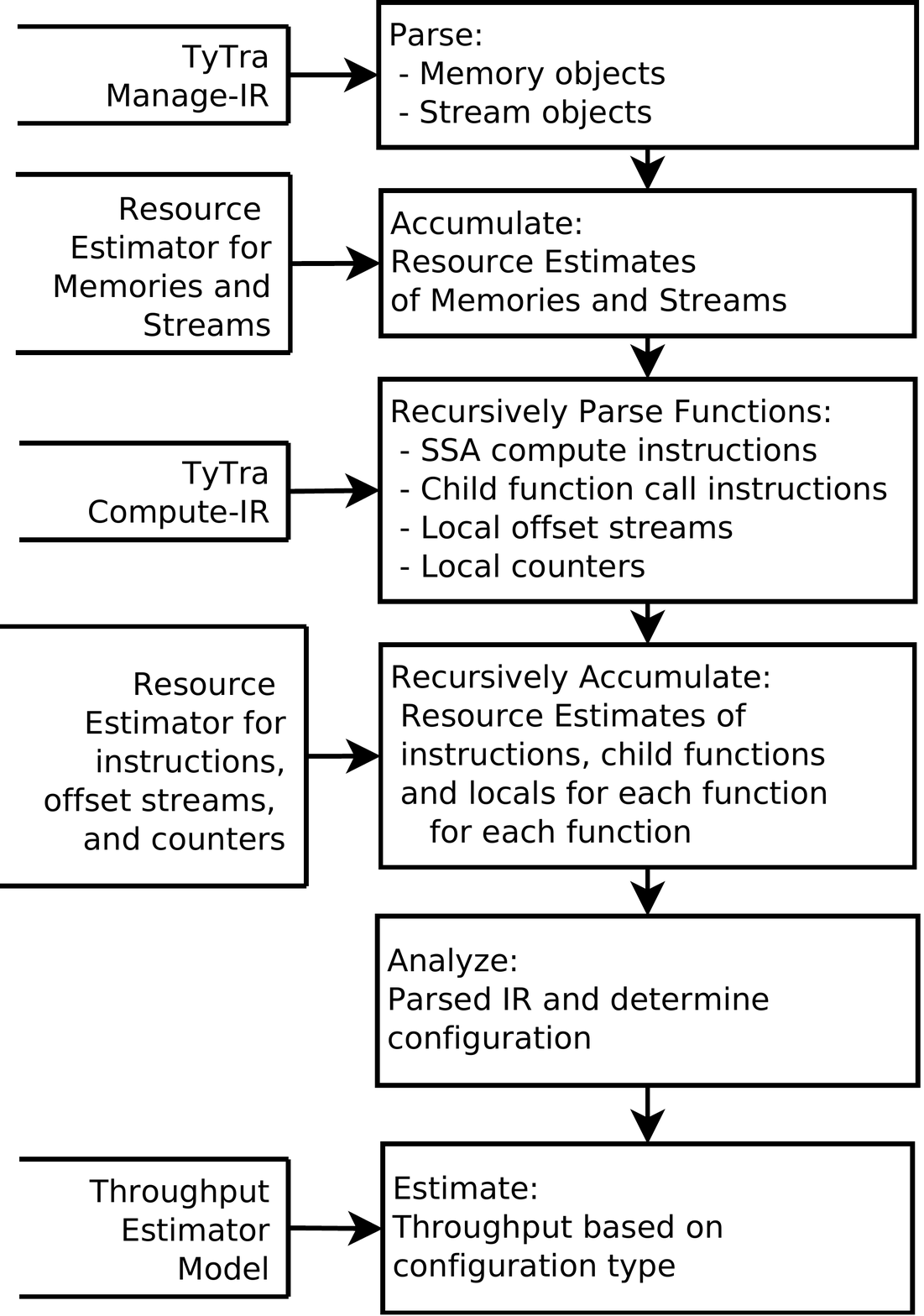}
\caption{TyTra Back-end Compliler's Flow for Generating Estimates}
\label{fig:TyTraEstimationFlow}
\end{figure}

We have written a TyTra Back-end Compiler (TyBEC) that generates estimates as described in this section. Figure \ref{fig:TyTraEstimationFlow} shows the flow of the TyBEC.  

Using the illustration from \S\ref{sec:TIR-illustration} we compared the estimates generated by TyBEC, with the actual resource consumption figures from synthesis of hand-crafted HDL. We only compare the two more relevant example for an FPGA, that is, a single pipeline configuration, and one where pipeline is replicated four times (\texttt{C2} and \texttt{C1}). The results of comparison are in Table \ref{tab:estVactualSimple}. Note that the purpose of these estimates primarily is to choose between different configurations of a kernel. The estimates are quite accurate and well within the tolerance set by the requirements.

\begin{table}[thb]
	\begin{tabular}{|p{14ex}|p{8ex}|p{8ex}|p{8ex}|p{8ex}|}
		\hline \textbf{Parameter}& \textbf{C2(E)} & \textbf{C2(A)} & \textbf{C1(E)} & \textbf{C1(A)} \\ 
		\hline \hline ALUTs 			& 82 	& 83 	&  36.3K& 37.6K \\ 
		\hline REGs 			& 172   & 177 	&  18.6K& 19.1K \\ 
		\hline BRAM(bits) 		& 7.20K & 7.27K &  216K & 221K	\\ 
		\hline DSPs 			& 1  	& 1     &  4  	& 4 	\\ 
		\hline
		\hline Cycles/Kernel	& 1003  & 1008 	& 250 	&  258	\\ 		
		\hline EWGT       		& 249K  & 292K  &  997K & 826K 	\\ 
		\hline 
	\end{tabular} 
	\caption{Estimated (E) vs actual (A) cost and throughput for \texttt{C2} and \texttt{C1} configurations of a very simple kernel}
	\label{tab:estVactualSimple}
\end{table}

\section{Case Study - Successive Relaxation}
\label{sec:caseStudy}

We discuss a more realistic kernel in this section, to demonstrate the expressibility of the TIR  and effectiveness of its cost model. The \textit{successive over-relaxation method} is a way of solving a linear system of equations, and requires taking a weighted average of neighbouring elements over successive iterations\footnote{The TIR has the semantics for standard and custom floating-point representation but the compiler does not yet support floats.}. The listing in Figure \ref{fig:code4paperRelax} is a C-style pseudo-code of the algorithm:

\begin{figure}[th]
\centering
\includegraphics[width=1.0\linewidth, frame]{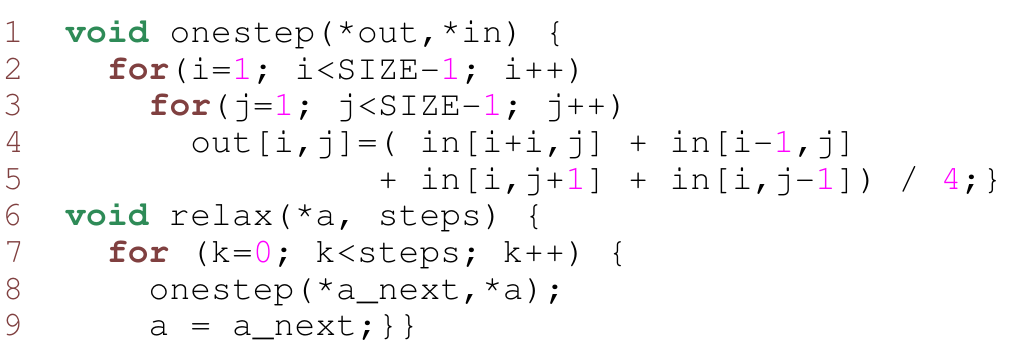}
\caption{C code for the successive relaxation algorithm}
\label{fig:code4paperRelax}
\end{figure}

\begin{figure}[th]
\centering
\includegraphics[width=0.9\linewidth, frame]{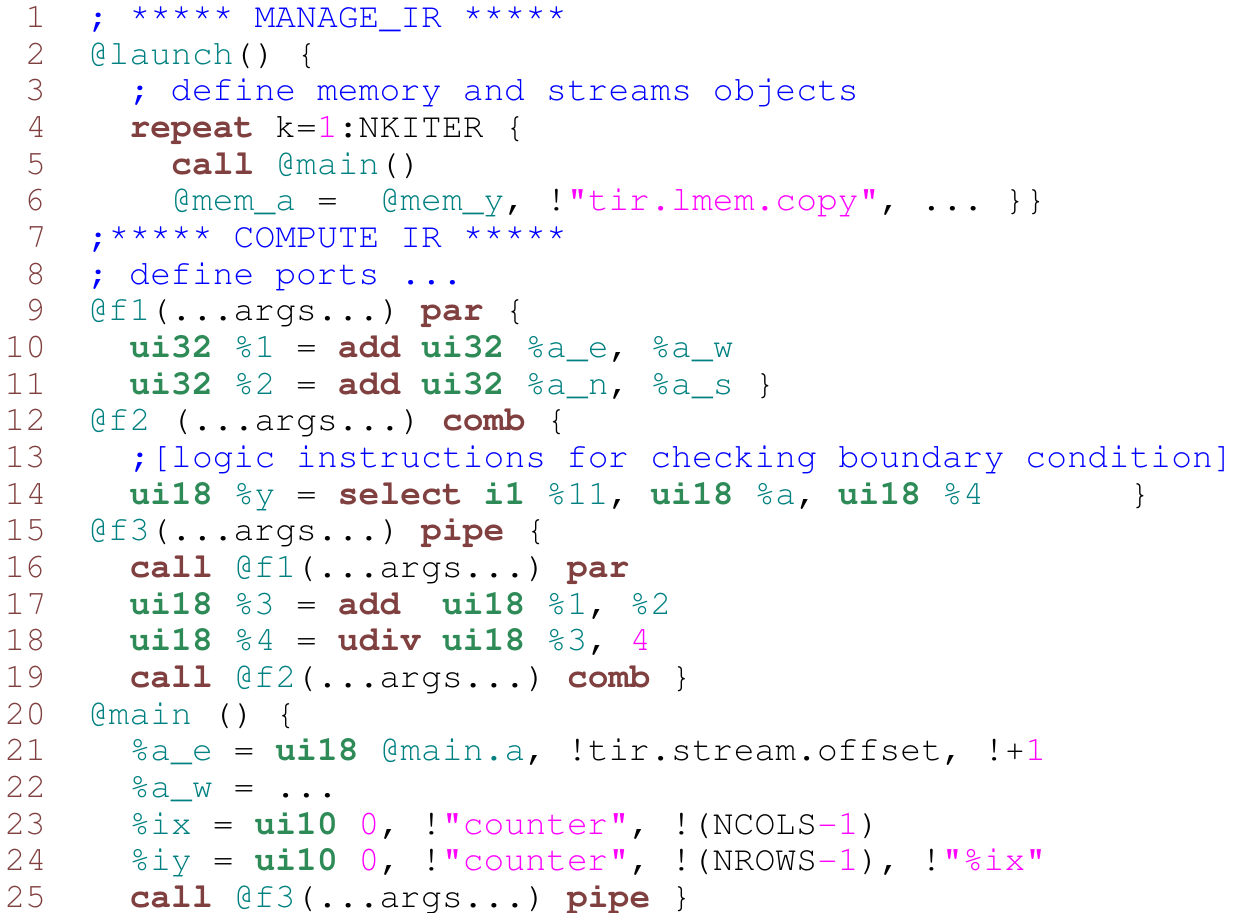}
\caption{TyTra-IR code for the relaxation kernel configured as a single pipeline}
\label{fig:code4paperRelaxC2}
\end{figure}

Figure \ref{fig:code4paperRelaxC2} shows how this translates to TyTra-IR configured as a single pipeline (\texttt{C2}). Note the use of stream offsets (line 21), repeated call to kernel through the \texttt{repeat} keyword (line 4), and use of a function of type \texttt{comb} (line 12), which translates to a single-cycle combinatorial block. We also use nested counters for indexing the 2D index-space (lines 23-24). 

We also implemented this kernel as another configuration with replicated pipelines (\texttt{C1}, similar to the configuration in \S\ref{sec:illustration-config3}).

\subsection*{Results of Estimator}
We ran the TyBEC estimator on the two configurations and compared the resource and throughput figures obtained from hand-crafted HDL. Table \ref{tab:estVactualRelax} shows this comparison.

\begin{table}
\begin{tabular}{|p{14ex}|p{8ex}|p{8ex}|p{8ex}|p{8ex}|}
	\hline \textbf{Resource}& \textbf{C2(E)} & \textbf{C2(A)} & \textbf{C1(E)} & \textbf{C1(A)} \\ 
	\hline ALUTs 			& 528 	& 546 	& 5764 	&  5837	\\ 
	\hline REGs 			& 534 	& 575 	& 4504 	&  4892 \\ 
	\hline BRAM(bits) 		& 5418 	& 5400 	& 11304 &  11250 \\ 
	\hline DSPs 			& 0 	& 0 	& 0 	&  0   \\ 
	\hline
	\hline Cycles/Kernel	& 292  	& 308 	& 180 	&  185	\\ 		
	\hline EWGT       		& 57K  	& 43K 	& 92K 	&  72K	\\ 		
	\hline 
\end{tabular} 
	\caption{Estimated (E) vs actual (A) cost and throughput for two configurations \texttt{C2} and \texttt{C1} of relaxation kernel}
	\label{tab:estVactualRelax}
\end{table}


A reasonable accuracy of the estimator is clearly indicated by these comparisons. This vindicates our observation that an IR designed at an appropriate abstraction will yield estimates of cost and performance in a very straightforward and light-weight manner, that are accurate enough to make design decisions. Hence it is our plan to use this IR to develop a compiler that takes legacy code, and automatically compares various possible configurations on the FPGA to arrive at the best solution.

\section{Related Work}
\label{sec:relatedWork}
High-Level Synthesis for FPGAs is an established technology both in the academia and research. There are two ways of comparing our work with others. If we look at the entire TyTra flow as shown in Figure \ref{fig:TyTra_Flow}, then the comparison would be against other C-to-gates tools that can work with legacy code and generate FPGA implementation code from it. As an example, LegUP\cite{112.153} is an upcoming tool developed in the academia for this purpose. Our own front-end compiler is a work in progress and is not the focus of this paper. 


A more appropriate comparison for this paper would be to take the TyTra-IR as a custom language that allows one to program FPGAs at a higher abstraction than HDL, and could be used as an automated or manual route to FPGA programming. Reference \cite{123.024} for example discussed the Chimps language that is at a similar abstraction as TIR and generates HDL description. Our work is relatively less developed compared to Chimps, however there is nothing equivalent to the estimator model that we have in the TyBEC. The MaxJ is a Java-based custom language used to program Maxeler DFEs (FPGAs) \cite{112.154}. It is in some ways similar to TIR in the way uses stream abstractions for data, creates pipelines by default. In fact, our IR has been informed a study of the MaxJ language. The use of streaming and scalar ports, offset streams, nested counters, and separation of management and computation code in the TIR is very similar to MaxJ. However, the similarity does not extend much beyond these elements. TIR and MaxJ are at very different abstraction levels, with the latter positioned to provide a programmer-friendly way to program FPGAs. %
The TIR on the other hand is meant to be a target language for a front-end compiler, and is therefore lower abstraction and fine-grained. This fine-grained nature allows a much better observability and controllability of the configuration on the FPGA, which makes it a more suitable language to explore the entire FPGA design space.

Altera-OCL is an OpenCL compatible development environment for targeting Altera FPGAs\cite{112.151}. It offers a familiar development eco-system to programmers already used to programming GPUs and many/multi-cores using OpenCL. A comparison of a high-level language like OpenCL with TyTra-IR would come to similar conclusions as arrived in relation to MaxJ. In addition, we feel that the intrinsic parallelism model of OpenCL, which is based on multi-threaded work-items, is not suitable for FPGA targets which offer the best performance via the use of deep, custom pipelines. 
%
Altera-OCL is however of considerable importance to our work, as we do not plan to develop our own host-API, or the board-package for dealing with FPGA peripheral functionality. We will wrap our custom HDL inside an OpenCL device abstraction, and will use OpenCL API calls for launching kernels and all host-device interactions.

\section{Conclusion and Future Work}

In this paper, we showed that the TIR syntax makes it very easy to construct a variety of configurations on the FPGA. While  the current semantics are already reasonably expressive,  the TIR will evolve to encompass more complex scientific kernels. We showed that we can estimate very accurate estimates of cost and performance from the TIR without any further translations. We indicated that automatic HDL generation is a straightforward process, which is a work in progress, and our immediate next step.

We plan to improve the accuracy of the estimator with better mathematical models. We will also  make our estimator tool more generic, as for this proof-of-concept work the supported set of instructions and data-types is quite limited. The compiler will also be extended to incorporate optimizations, in particular we aim to incorporate LegUP's  sophisticated LLVM  optimizations before emitting HDL code\cite{112.153}. 


While we work on maturing the TIR and the back-end compiler, we will be moving higher up the abstraction as well. We will be investigating the automatic generation of the TIR from HLL, with the help of Multi-Party Session Types to ensure correctness of transformations. 

\section*{Acknowledgement}
The authors acknowledge the support of the EPSRC for the TyTra project (EP/L00058X/1).

\bibliography{../bibUniversalWN} 
\bibliographystyle{IEEEtran} 

\end{document}